\documentclass[preprint,preprintnumbers,amsmath,amssymb]{revtex4}
\usepackage{amsfonts}
\usepackage{amssymb}
\usepackage{latexsym}
\usepackage{graphicx}% Include figure files
\newcommand{\al}{\alpha}

\newcommand{\be}{\beta}

\newcommand{\si}{\sigma}
\newcommand{\Si}{\Sigma}

\newcommand{\De}{\Delta}

\begin{document}

\title{The $HeH^+$ molecular ion in a magnetic field.}

\author{A.~V.~Turbiner}
\email{turbiner@nucleares.unam.mx}
\author{N.~L.~Guevara}
\email{nicolais@nucleares.unam.mx}
\affiliation{Instituto de Ciencias Nucleares, Universidad Nacional
Aut\'onoma de M\'exico, Apartado Postal 70-543, 04510 M\'exico,
D.F., Mexico}
\date{\today}

\begin{abstract}
A detailed study of the low-lying electronic states
${}^1\Si,{}^3\Si,{}^3\Pi,{}^3\De$  of the $\rm{HeH}^+$ molecular ion
in parallel to a magnetic field configuration (when $\al$-particle
and proton are situated on the same magnetic line) is carried out
for $B=0-4.414\times 10^{13}$\,G in the Born-Oppenheimer
approximation. The variational method is employed using a physically
adequate trial function. It is shown that the parallel configuration
is stable with respect to small deviations for $\Si$-states. The
quantum numbers of the ground state depend on the magnetic field
strength. The ground state evolves from the spin-singlet ${}^1\Si$
state for small magnetic fields $B\lesssim 0.5 $\,a.u. to the
spin-triplet ${}^3\Si$ unbound state for intermediate fields and to
the spin-triplet strongly bound $^3\Pi$ state for $B \gtrsim 15
$\,a.u. When the $\rm{HeH}^+$ molecular ion exists, it is stable
with respect to a dissociation.
\end{abstract}

%\pacs{31.15.Pf,31.10.+z,32.60.+i,97.10.Ld}

\maketitle

\section{\protect\bigskip introduction}

It is well-known that a very strong magnetic field may appear on the
surface of neutron stars, $B\approx 10^{11}-10^{13}$\,G. It seems
natural to expect that in a vicinity of the surface some atomic and
molecular systems can occur. For many years this assumption has
motivated a development of atomic-molecular physics in a strong
magnetic field hoping that it might lead to understanding of the
spectra of these compact stars and be important for a construction
of a model of the atmosphere. Recent revolutionary observation of
the absorption features in a spectra of the isolated neutron star
1E1207.4-5209 by Chandra $X$-ray observatory \cite{Sanwal:2002}
brought an extra attention to the field.

So far, a major attention has been paid to a study of one-electron
atomic and molecular systems (for a review, see \cite{PR}). During
the last years the existence of several exotic, strongly bound
one-electron molecular systems has been predicted. In particular, it
was shown that for a sufficiently strong magnetic field $B\gtrsim
5\times 10^{12}\,\mbox{G}$ the exotic ion $(He H)^{2+}$ can exist in
parallel configuration (the magnetic field is directed along
internuclear axis) as optimal as well as its excited states $1\pi,
1\delta$ \cite{HeH++}. A complete list of possible one-electron
atomic and molecular systems is known for $B \lesssim
B_{schwinger}=4.414 \times 10^{13}$\,G. In turn, the exploration of
two-electron systems has been mostly restricted to a study of
atomic-type systems H$^-,\ $He with the only exception of the H$_2$
molecule (see \cite{turbinerH2}, \cite {schmelcherH2} and references
therein). Recently, a first detailed study of the molecular ions
H$_3^+$ \cite{Turbiner:2006H3+} and $\rm{He}_2^{2+}$
\cite{Turbiner:2006He2++} has been carried out (for a brief review
of a general situation see \cite{Turbiner:2006London}). It was shown
that for large magnetic fields if such an ion exists its optimal
configuration is the linear parallel one - heavy charged particles
are situated along a magnetic line.
%As a magnetic field increases some surprising evolution of the
%ground state quantum numbers take place.
A goal of the present paper is to make a study of the Coulomb system
$(\al p e e)$ in a magnetic field ranging in $B=0-4.414 \times
10^{13}\,$\,G and thus to establish the possible existence of the
simplest two-electron diatomic heteronuclear molecular ion
$\rm{HeH}^{+}$ and to explore its lowest excited states. In the
field-free case the theoretical existence of $\rm{HeH}^{+}$ was
established long ago by Beach \cite{beach} (see also \cite{anex})
while its several excited states were found (see, e.g.,
\cite{3sigma}). Experimentally, this molecular ion was first
observed in the mass spectra of helium-hydrogen mixtures (see, e.g.,
\cite{hogness}).
%Studies of $\rm{HeH}^+$ seem quite relevant from astrophysical point
%of view, since protons and $\al$-particles together with electrons
%are the most abundant elements in Universe \cite{dalgarno}.
The $\rm{HeH}^+$ ion is thought to be among the first molecular
species to appear in the Universe (see \cite{lepp} and references
therein). An attempt to explore the $\rm{HeH}^{+}$ ion in
ultra-strong magnetic field was made in \cite{heyl}.

Atomic units are used throughout ($\hbar$=$m_e$=$e$=1), although
energies are expressed in Rydbergs (Ry). The magnetic field $B$ is
given in a.u. with $B_0= 2.35 \times 10^9\,G$.

\section{Generalities}

We study the system $(\al p e e)$ placed  in a uniform constant
magnetic field. It is assumed that the $\al$-particle and proton are
infinitely massive (Born-Oppenheimer approximation of zero order).
It seems natural from physical point of view that the optimal
configuration is parallel when the $\al$-particle and proton are
situated along the magnetic field line. Nevertheless a stability of
this configuration with respect to small perturbations will be
checked.

\begin{figure}[tb]
\begin{center}
   \includegraphics*[width=5.in,angle=0.0]{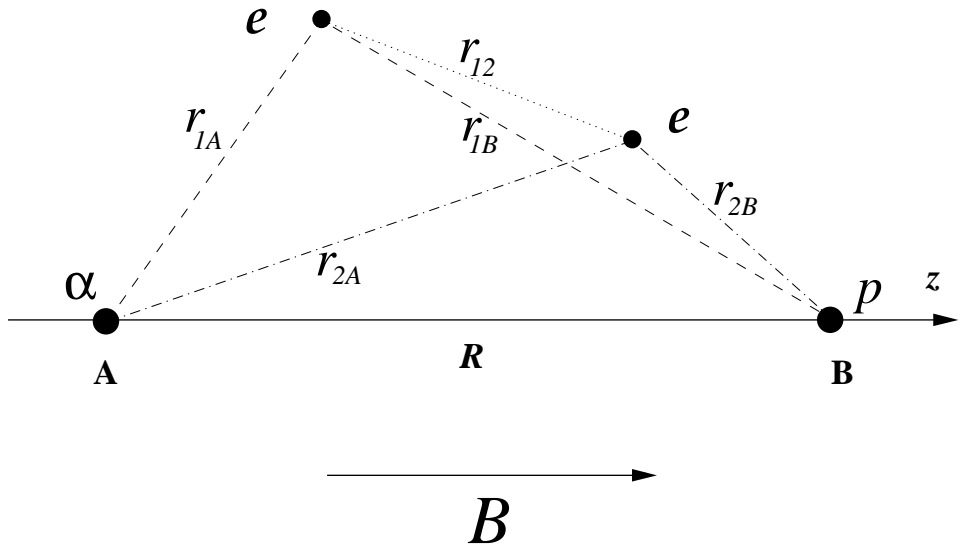}
    \caption{\label{fig:h3pp} Geometrical setting  for the $\rm{HeH}^+$
      ion parallel to a magnetic field directed along the $z$-axis. The
     $\al$-particle and the proton (marked by bullets) are situated on the $z$-line at
      distance $R$.}
\end{center}
\end{figure}

The Hamiltonian which describes the system $(\al p e e)$ when the
magnetic field is oriented along the $z$ direction, ${\bf
B}=(0,0,B)$ is
\begin{equation}
 {\cal H} =\sum_{\ell=1}^2 \left( {\hat {\mathbf p}_{\ell}+{\cal A}_{\ell}}
 \right)^2 -\sum_{\buildrel{{\ell}=1,2}\over{\kappa =\al,p}}
 \frac{2Z_{\kappa}}{r_{{\ell},\kappa}}
 + \frac{2}{r_{12}}+ \frac{4}{R} + 2{\bf{B}} \cdot {\bf{S}}\  ,
\end{equation}
(see Fig.~\ref{fig:h3pp} for the geometrical setting and notations),
where ${\hat {\mathbf p}_{\ell}}=-i \nabla_{\ell}$ is the momentum
of the ${\ell}$-th-electron, the index $\kappa$ runs over the
$\al$-particle and proton, $Z_{\al}=2, Z_p=1$ and $r_{12}$ is the
interelectronic distance, $\bf{S}=\hat s_{1}+\hat s_{2}$ is the
operator of the total spin. The vector potential ${\cal A}_{\ell}$
which corresponds to the constant uniform magnetic field $\bf B$ is
chosen in the symmetric gauge,
\begin{equation}
   {\cal A}_{\ell}= \frac{1}{2}({\bf{B}} \times \ {\bf{r}}_{\ell})=
   \frac{B}{2} (-y_{\ell},\ x_{\ell},\ 0)\ .
\end{equation}

%%%%%%%%%%%%%%  FIGURE:1  %%%%
%%%%%%%%%%%%%%  H3++ geometrical settings

\noindent Finally, the Hamiltonian can be written as
\begin{equation}
 %\hspace{-10pt}
  {\cal H} =\sum_{{\ell}=1}^2 \left(- {\mathbf\nabla}^2_ {\ell}
  +\frac{B^2}{4} \rho_{\ell}^2 \right) -
  \sum_{\buildrel{{\ell}=1,2}\over{\kappa =\al,p}}
  \frac{2Z_{\kappa}}{r_{{\ell}\kappa}}  + \frac{2}{r_{12}}+
  \frac{4}{R} + B (\hat L_z +2\hat S_z )\  ,
\end{equation}
where  $\hat L_z=\hat l_{z_1}+\hat l_{z_2}$ and $\hat S_z=\hat
s_{z_1}+\hat s_{z_2}$ are the z-components of the total angular
momentum and total spin, respectively, and $\rho_{\ell}^2 =
{x_{\ell}^2+y_{\ell}^2}\,,\ \ell =1,2$.

% The conserved quantum numbers
%are the eigenvalues of \hat S_z $ and $\hat L_z  $, and
%parity  $\hat \Pi $ (alternatively, we can use the z-parity, $\hat
%P $ ). The quantum numbers associated with these operators are
%denoted respectively $M_S$, $M_L$ and $(-1)^{P}$ (with $P$ taking
%the values 0 and 1).

%%%%%%%
The problem is characterized by two conserved quantities: (i) the
operator of the $z$-component of the total angular momentum
(projection of the angular momentum on the magnetic field direction)
giving rise to the magnetic quantum number $m$ and (ii) the operator
of the $z$-component of the total spin (projection of the total spin
on the magnetic field direction) giving rise to the spin quantum
number $m_s$. Hence, any eigenstate has two quantum numbers
assigned: the magnetic quantum number $m$ and the spin quantum
number $m_s$.

As the magnetic field increases a contribution from the linear
Zeeman term (interaction of the spin with a magnetic field ${\bf{B}}
\cdot {\bf{S}}$) becomes more and more important. Hence, it is
natural to assume that for large magnetic fields the states with
minimal $m_s$ (when both electron spins are antiparallel to the
magnetic field) will represent the states with lowest total
energies.

The space of eigenstates is split into subspaces (sectors) with each
of them characterized by definite values of $m$ and $m_s$. Thus, to
classify eigenstates we follow the convention widely accepted in
molecular physics using the quantum numbers $m$ and $m_s$. The
notation is ${}^{2S+1} M$, where $2S+1$ is the spin multiplicity and
corresponds to $1$ for singlet $S=0$ and  $3$ for triplet $S=1$, as
for the labeling $M$ it is used the Greek letters $\Si, \Pi, \De$
that correspond to the states with $|m|=0, 1, 2,...$, respectively.
We study only states with $m=0,-1,-2$.

As a method to explore the problem we use the variational procedure.
The recipe of choice of trial functions is based on physical
arguments \cite{turbinervar}. A trial function for states with
orbital magnetic quantum number $m$ is chosen in the form
\begin{equation}
\label{ansatz}
  \psi^{(trial)} =(1+\sigma_e P_{12})
  \rho_1^{\mid m \mid}e^{im\phi_1}
  {e}^{-\al_1 r_{1A}-\al_2 r_{1B} -\al_3 r_{2A} -
  \al_4 r_{2B} - B  \be_{1}\rho_1^2/4-
  B \be_{2}\rho_2^2/4+ \gamma r_{12} }\ ,
\end{equation}
(cf.\cite{turbinerH2}), where $\sigma_e=1,-1$ stand for spin singlet
($S=0$) and triplet states $(S=1)$, respectively. The $P_{12}$ is
the operator which interchanges electrons (1 $\leftrightarrow$ 2).
The parameters $\al_{1-4}$, $\be_{1-2}$ and $ \gamma$ as well as the
internuclear distance $R$ are eight variational parameters.

Calculations were performed using the minimization package MINUIT
from CERN-LIB. Multidimensional integration was carried out using a
dynamical partitioning procedure: a domain of integration was
divided into subdomains following the integrand profile and then
each subdomain was integrated separately (for details see, e.g.,
\cite{PR}). Numerical integration was done with a relative accuracy
of $\sim 10^{-6} - 10^{-7}$ by use of the adaptive D01FCF routine
from NAG-LIB. Computations were performed on a dual DELL PC with two
Xeon processors of 2.8\,GHz each (ICN) and dual DELL PC with two
Xeon processors of 3.06\,GHz each (CINVESTAV). We implemented a
parallelization in the routine for the calculation of the integrals.
Minimization process for each particular magnetic field and for a
given state was quite painful and lengthy, it took from 100 to 1000
hours of CPU time, however, when a minimum is found it was needed a
few minutes to compute a minimal variational energy.

\section{Results}
%%%%%%%%%%%%%%%%%%%%%%%%%
In absence of a magnetic field the $\rm{HeH}^+$ molecular ion exists
and it is stable. Its ground state is the ${}^1\Si$ ($S = 0$ and
$m=0$) state \cite{beach}. In order to establish the existence of
$\rm{HeH}^+$ in a magnetic field we carried out a detailed study of
the Coulomb system $(\al p e e)$ in parallel configuration in the
state ${}^1\Si$ for magnetic fields $B=0-10000\,$ a.u. (see Table
\ref{table1}). For this case the variational trial function
$\psi^{trial}$ (4) is used at $\si_e =1$ and $m=0$.

The calculations indicate the optimal geometry is parallel
($\al$-particle and proton stay on the same magnetic line). For this
configuration for all studied magnetic fields a minimum in the total
energy curve $E_T(R)$ is developed. Furthermore, we checked that
this minimum is stable with respect to small inclinations of the
internuclear axis from the magnetic field direction. It is found
that with an increase of the magnetic field strength the total
energy of the $(\al p e e)$ system in the ${}^1\Si$ state grows, it
becomes more bound (in particular, the double ionization energy
increases $E_I=2B-E_T$) and the system becomes more compact (the
internuclear equilibrium distance decreases), see Table~I. For
$B\leq 100$\,a.u. the $(\al p e e)$ system in the ${}^1\Si$ state is
stable with respect to the dissociation $\rm{He^+}$ + H or $\rm{He}$
+ p. The vibrational and rotational energies are calculated for
$B=0.5$\,a.u. The vibrational energy is larger than the rotational
energy which is typical for field-free case.

\begin{table}
  \centering
\caption{
\label{table1}
 $\rm{HeH}^+$ ion in the state  $^1\Si$:
 Total $E_T$, double ionization $E_I$, vibrational $E_0^{vib}$ and rotational
 $E_0^{rot}$ energies in Ry, and equilibrium distance in a.u. of
 the $(\al p e e)$ system in the state $^1\Si$. Magnetic fields for which
 the $^1\Si$ state is the ground state are marked by ${}^{*}$ (see text).
 ${}^a$ Result from \cite{beach} (rounded).
 Total energy $ E(*)= E(\rm{He}^+(1s))+E(\rm{H}(1s))$
 in Ry with electrons in spin-singlet state is from \cite{PR}.
 Total energy of He are from \cite{schmelcherHe1}, $^{\times}$ marks a result
 of linear interpolation.
 }
\begin{tabular}{c|c|c|c|c|c|c|c}
\hline\hline
  B (a.u)      & $E_T$ & $E_I$ & $R_{eq}$ & $E_0^{vib}$ & $E_0^{rot}$
                                      & $E(*)$ & $ E(He(1^10^+))$  \\
\tableline
$0^{*}$     & -5.943      & 5.943  & 1.451   &\,0.016 &        & -5.0    & -5.807 \\
            & -5.957 $^a$ &        & 1.463   &        &        &         &        \\
$0.5^{*}$   & -5.843      & 6.843  & 1.430   &\, 0.018&\,0.005 &
-4.8$^{\times}$ & -5.6$^{\times}$ \\
            &             &        &         &        &        &          &        \\
1           & -5.626      & 7.626  & 1.404   &        &        & -4.544   & -5.459 \\
            &             &        &         &        &        &          &        \\
10          &  5.337      & 14.663 & 0.789   &        &        &  7.725   &  6.129  \\
            &             &        &         &        &        &          &        \\
100         & 166.513     & 33.487 & 0.370   &        &        & 173.327  & 169.837 \\
            &             &        &         &        &        &          &        \\
1000        & 1927.35     & 72.65  & 0.212   &        &        & 1944.37  &        \\
            &             &        &         &        &        &          &        \\
10000       & 19848.67    & 151.30\, & 0.0928\ &       &        & 19893.62 &        \\
            &             &        &         &        &        &          &      \\
\tableline   \hline
\end{tabular}
\end{table}

We carried out a detailed study of the spin-triplet state ${}^3\Si$
($S = 1$ and $m=0$) of the $(\al p e e)$ system in parallel
configuration in the domain of magnetic fields $0 \le B \le
10000\,$a.u. For this state the variational trial function
$\psi^{trial}$ (4) with $\si_e =-1$ and $m=0$ is used.

Similar to ${}^1\Si$ all total energy curves $E_T(R)$ of the $(\al p
e e)$ system for magnetic fields $B=0-10000$\,a.u. in the ${}^3\Si$
state displays a minimum. However, to the contrary to the ${}^1\Si$
state the minimum is always shallow, see Fig.~2-5 as an
illustration. It is found that with an increase of the magnetic
field strength the total energy decreases. The system gets more
"compact" - the internuclear equilibrium distance of a shallow
minimum decreases slowly, see Table~\ref{table2}. We expect that the
system in the ${}^3\Si$ state "exists" in a form of two separated
atomic-like systems $\rm{He}^+(1s)$ + H$(1s)$ situated at large
distances with electron spins directed against a magnetic field
direction.

A comparison of the total energies of the $(\al p e e)$ system in
${}^1\Si$ and ${}^3\Si$ states in equilibrium indicates to the
appearance of a level crossing (see Tables \ref{table1} and
\ref{table2}). This crossing of the ${}^1\Si$ and  ${}^3\Si$ states
occurs at $B \approx 0.5 $\,a.u. It implies that the ${}^1\Si$ state
is the lowest energy state at $B \lesssim 0.5 $\,a.u., where the
$(\al p e e)$ system is stable towards possible decays. Hence, one
can conclude that the stable $\rm{HeH}^+$ molecular ion exists at $B
\lesssim 0.5 $\,a.u with ${}^1\Si$ state as the ground state. For $B
\gtrsim 0.5$\,a.u. the situation is changed - the ${}^3\Si$ state
may become the ground state of the $\rm{HeH}^+$ molecular ion. In
fact, the state ${}^3\Si$ is at most metastable with very short
lifetime or very likely is simply unbound - the $\rm{HeH}^+$ ion
dissociates to $\rm{He}^+(1s) + \rm{H}(1s)$ with the electrons in
spin-triplet state. A general conclusion can be drawn that for $0.5
< B < 15\,$a.u. the system $(\al p e e)$ likely exists in a form of
two separate atomic-type systems He$^+(1s) + \rm{H}(1s)$ with total
electron spin equal to one.

%%%%%%%%%%%%%%%%%%%%%%%%%%%%%%%%%%%%
\begin{table}
  \centering
\caption{
\label{table2}
 $\rm{HeH}^+$ ion in the state  ${}^3\Si$:
 Total $E_T$, double ionization $E_I$ energies in Ry and equilibrium distance in
 a.u. of the $(\al p e e)$ system in the state $^1\Si$.
 Magnetic fields for which the ${}^3\Si$ state might be the ground state
 are marked by ${}^{*}$. Total energy $E(*)= E(\rm{He}^+(1s))+E(\rm{H}(1s))$ (in Ry)
 for the electrons in spin-triplet state is from \cite{PR}. Total
 energies for the He atom with total electron spin 1 are
 from \cite{schmelcherHe1}. ${}^a$ Ref \cite{3sigma}. }
\begin{tabular}{c|c|c|c|c}
\hline
\hline
B(a.u)    &$E_T$ &$R_{eq}$   &  $ E(*)$ & $E(He)(1^3 0^+)$ \\
\tableline
0         & -5.0064  &\ 4.50 \  & -5.0    & -4.3504 \\
          &-5.00878$^a$&\ 4.35\ &         &     \\
$0.5^{*}$ & -5.8712  &\ 4.32 \  & -5.8637 & -5.3796$^{**}$  \\
          &          &        &         &     \\
$1^{*}$   & -6.5524  &\ 4.04 \  & -6.5443 & -5.3013 \\
          &          &        &         &  \\
$10^{*}$  & -12.209  &\ 2.70 \  & -12.275 & -9.255  \\
          &          &        &         &   \\
$15^{*}$  & -13.936  &\ 2.47 \  & -14.043 & -10.622$^{**}$ \\
          &          &        &         &   \\
%20       & -15.327  &\ 2.321 \ &          &         \\
%         &          &        &          &   \\
100       & -26.172  &\ 1.68 \  &  -26.673 & -19.686 \\
          &          &        &          &  \\
1000      & -54.020  &\ 1.20 \ &  -55.628 &    \\
          &          &        &          &  \\
10000     & -103.07  &\ 0.83 \  & -106.38 &   \\
          &          &        &         &  \\
\tableline          \hline
\end{tabular}
\\
\end{table}

As next we carried out a detailed study of the state ${}^3\Pi$ ($S =
1$ and $m=-1$) of the $(\al p e e)$ system in parallel configuration
for magnetic fields $0 \le B \le 4.414\times 10^{13}$\,G (see
Figs.~2-5). For this state the same variational trial function
$\psi^{trial}$ (4) is used at $\si_e =-1$ and $m=-1$.

The results of the calculations indicate unambiguously the existence
of a deep minimum in the total energy curve $E_T(R)$ of the $(\al p
e e)$ system in parallel configuration for all studied magnetic
fields $0 \le B \le 4.414\times 10^{13}$\,G. Hence, it can be
concluded that the $\rm{HeH}^+$ molecular ion may exist in the
${}^3\Pi$ state. Table~\ref{table3} contains the results for the
total energy $E_T$ and the internuclear equilibrium distance
$R_{eq}$ for the state ${}^3\Pi$. Perhaps, it is worth mentioning
that the previously obtained results for ultra-magnetic fields
\cite{heyl} are less accurate for both studied two-electron systems:
(i) the $\rm{HeH}^+$ ion in the present calculation and (ii) the
$\rm{He}$ atom in the calculation made in \cite{schmelcherHe}. With
an increase of the magnetic field strength the total energy
decreases, the system becomes more bound (double ionization energy
increases being equal for triplet states to $E_I=-E_T$) and more
compact (the internuclear equilibrium distance decreases).
Table~\ref{table3} shows the total energies of three dissociation
channels of $\rm{HeH}^+$ in ${}^3\Pi$ state: $\to \rm{He}(1^3(-1)^+)
+ p$, $\to \rm{He}^+(1s) + \rm{H}(2p_{-1})$ and $ \to
\rm{He}^+(2p_{-1}) + \rm{H}(1s) $ for different magnetic fields. It
can be seen that the state $^3\Pi$ is stable towards any
dissociation. It is worth noting that the dissociation energy needed
increases monotonously in all range of studied magnetic fields.
Therefore the stability of the system with a magnetic field
increases.

One can see that the crossing of the total energies of the ${}^3\Pi$
and ${}^3\Si$ states at equilibrium occurs at $B \approx 15$\,a.u.
For larger magnetic fields the total energy of the ${}^3\Pi$ state
is lower than the total energy of ${}^3\Si$. Hence, the ground state
for the $\rm{HeH}^+$ ion for $B \gtrsim 15$\,a.u. is realized by the
${}^3\Pi$ state. Eventually, one can draw a conclusion that the
${}^3\Si$ state is the lowest energy state of the $(\al p e e)$
system for $0.5 < B \lesssim 15$\,a.u. and the $\rm{HeH}^+$ ion can
exist.

\begin{table}
  \centering
\caption{
\label{table3}
  $\rm{HeH}^+$ ion in the state  ${}^3\Pi$:
  Total $E_T$ and vibrational $E_{vib}$ energies in Ry and equilibrium distance
  $R_{eq}$ in a.u. of the $\rm{HeH}^+$ ion in the ${}^3\Pi$ state.
  Magnetic fields for which the ${}^3\Pi$ state is the ground state are
  marked by ${}^{*}$. Total energies $E(**)= E(\rm{He}^+(2p_{-1})+\rm{H}(1s))$
  and $ E(***)=  E(\rm{He}^+(1s)+\rm{H}(2p_{-1}))$ (in Ry) for the electrons
  in spin-triplet state from \cite{PR}.
  Total energy for He atom in spin-triplet electronic state in magnetic field
  is from \cite{schmelcherHe} (rounded).
  $^{\star}$ Result of interpolation.
  ${}^a$ from~\cite{3sigma}.
  ${}^b$ from~\cite{schmelcherHe1}.
  ${}^c$ from~\cite{heyl}.
 }
\begin{tabular}{c|c|c|c|c|c|c}
\hline\hline
B(a.u)      & $E_T$   & $R_{eq}$ & $E_{vib}$ & $E(**)$
& $E(***)$ &
            $E(He(1^3(-1)^+))$ \\
\tableline
0           &-4.263     & 7.61   &      & -2.0    &  -4.25   & -4.266$^b$ \\
            &-4.276$^a$& 7.7    &      &         &          &             \\
            &           &        &      &         &          &            \\
1           & -5.975    & 3.297  &      & -3.829  & -5.795   & -5.919  \\
            &           &        &      &         &          &         \\
10          & -12.026   & 1.135  &      & -8.732  & -11.030  & -11.659 \\
            &           &        &      &         &          &          \\
$15^{*}$    & -13.96    & 0.925  & 0.04 &         & -12.66   & -13.58$^{\star}$ \\
            &           &        &      &         &          &         \\
$100^{*}$   & -28.36    & 0.440  & 0.15 & -20.30  & -24.37   & -26.15  \\
            &           &        &      &         &          &         \\
$1000^{*}$  & -64.24    & 0.203  & 0.50 & -43.98  & -51.60   & -56.06  \\
            & -63.07$^c$ &       &      &         &          & -54.95$^c$ \\
$10000^{*}$ & -133.49   & 0.104  & 1.41 & -86.79  & -99.98   & -110.30 \\
            & -129.7$^c$ &       &      &         &          & -106.9$^c$ \\
$4.414 \times 10^{13}\,$G $^{*}$ & -160.50 & 0.092 &  &  &         &  \\
\tableline
\hline
\end{tabular}
\\
\end{table}

In order to complete the study we consider the state ${}^3\Delta$
($S = 1, m=-2$ ) of the $(\al p e e)$ system in the domain of
magnetic fields $0 \le B \le 4.414\times 10^{13}$\,G. It is worth
mentioning that the state $^3\Delta$ for the field-free case was
studied in \cite{3sigma} and \cite{Greennew}. For this state the
same variational trial function $\psi^{trial}$ (4) is used but with
$\sigma_e =-1$ and $m=-2 $. For all total energy curves $E_T(R)$ for
different magnetic fields the minimum is found. Hence, the ion
$\rm{HeH}^+$ may exist in ${}^3\Delta$ state for all magnetic
fields. Table~\ref{table4} shows the total $E_T$ energy and the
internuclear equilibrium distance $R_{eq}$ for the state
${}^3\Delta$. With an increase of the magnetic field strength the
total energies decrease, the system becomes more bound (double
ionization energies increase, $E_I=-E_T$, for triplet states) and
more compact (the internuclear equilibrium distance decreases). At
$B \approx 300$\,a.u. a crossing of the total energies of ${}^3\Si$
and ${}^3\Delta$ at equilibrium occurs. For larger magnetic fields
the total energy of the ${}^3\Delta$ is smaller than the ${}^3\Si$
state. It implies that for $B \gtrsim 300$\,a.u. the ${}^3\Delta$
state is the lowest excited state.

\begin{table}
  \centering
\caption{
\label{table4}
  Total energy in Ry and equilibrium distance in a.u.
  of the $\rm{HeH}^{+}$ ion in the state  ${}^3\Delta$.}
\begin{tabular}{c|c|c}
\hline\hline
B(a.u.)    & $E_T$         & $R_{eq}$ \\
\tableline
0          & -4.108        & 16.97  \\
1          & -5.664        & 4.717  \\
10         & -11.02        & 1.581  \\
100        & -25.78        & 0.486  \\
1000       & -58.68        & 0.218  \\
10000      & -123.62       & 0.112  \\
$4.414 \times 10^{13}\,$G  & -149.0 & 0.096\\
%          &            &           \\
\tableline\hline
\end{tabular}
\\
\end{table}

\begin{figure}[tb]
\begin{center}
   \includegraphics*[width=3in,angle=-90]{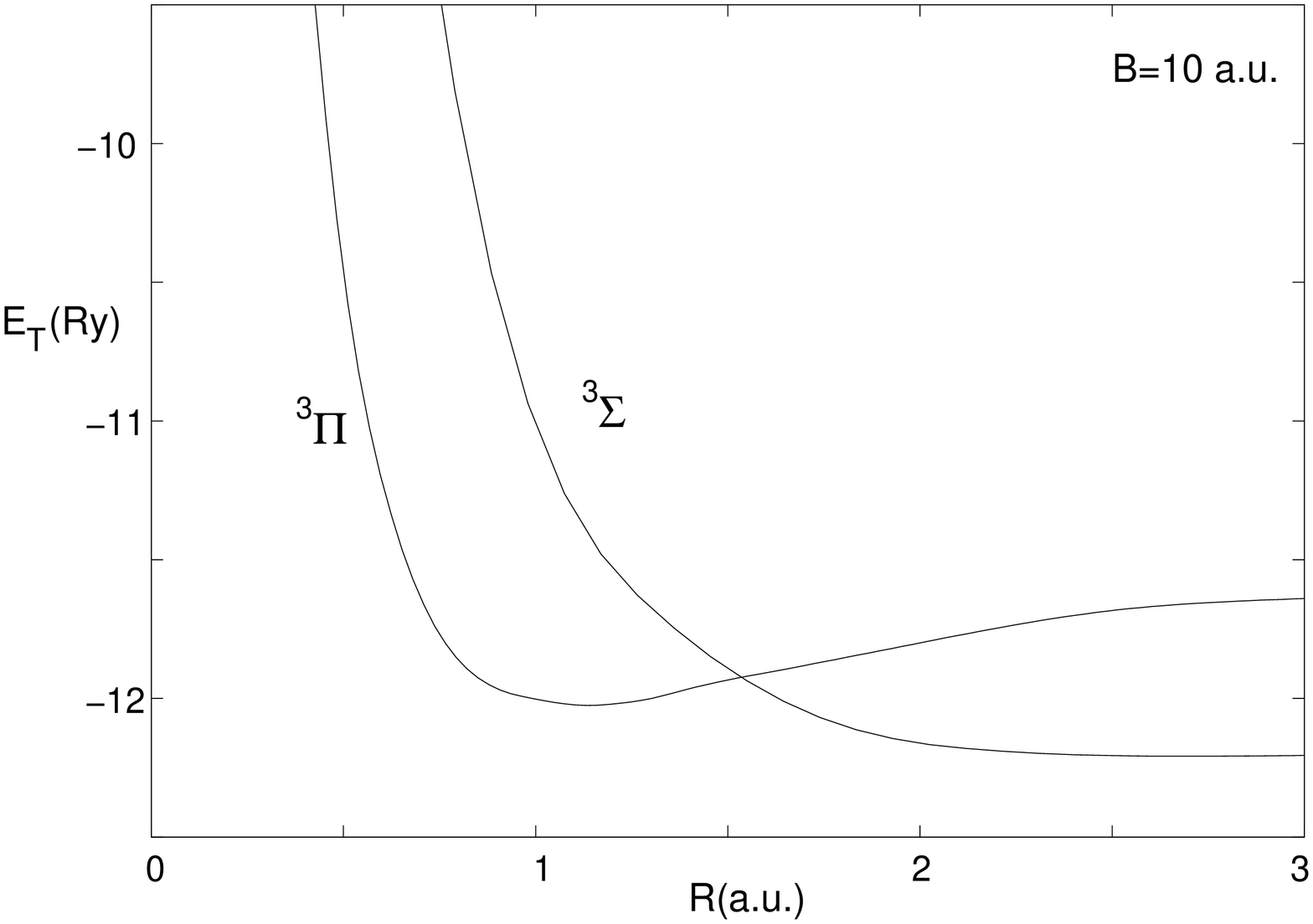}
    \caption{\label{fig:fig5} Total energy curves for triplet states
    $^3\Si$ and $^3\Pi$ for a magnetic field $B=10 $\,a.u.}
\end{center}
\end{figure}

\begin{figure}[tb]
\begin{center}
   \includegraphics*[width=3in,angle=-90]{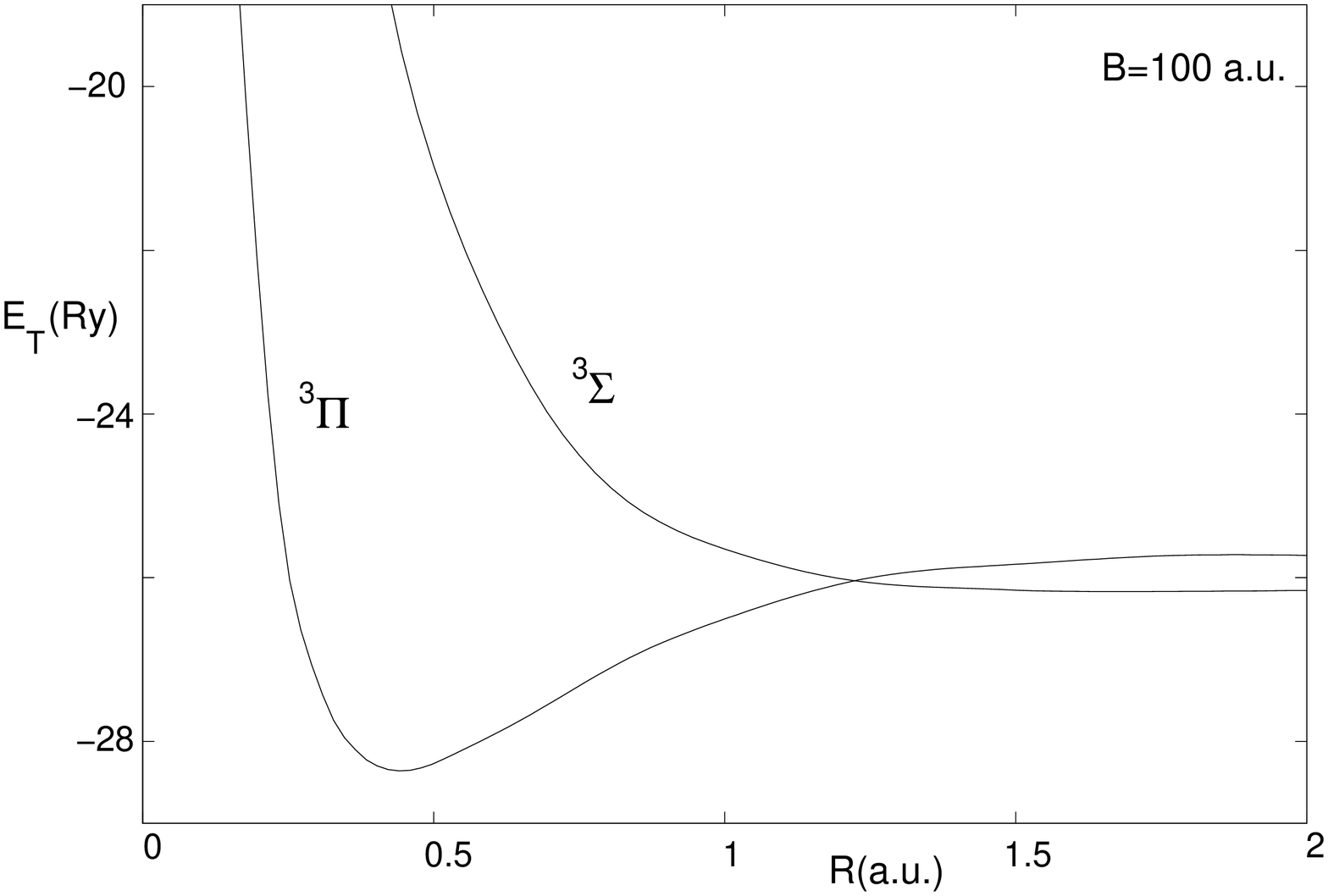}
    \caption{\label{fig:fig6} Total energy curves for triplet states
    $^3\Si$ and $^3\Pi$ for a magnetic field
    $B=100 $\,a.u.}
\end{center}
\end{figure}

\begin{figure}[tb]
\begin{center}
   \includegraphics*[width=3in,angle=-90]{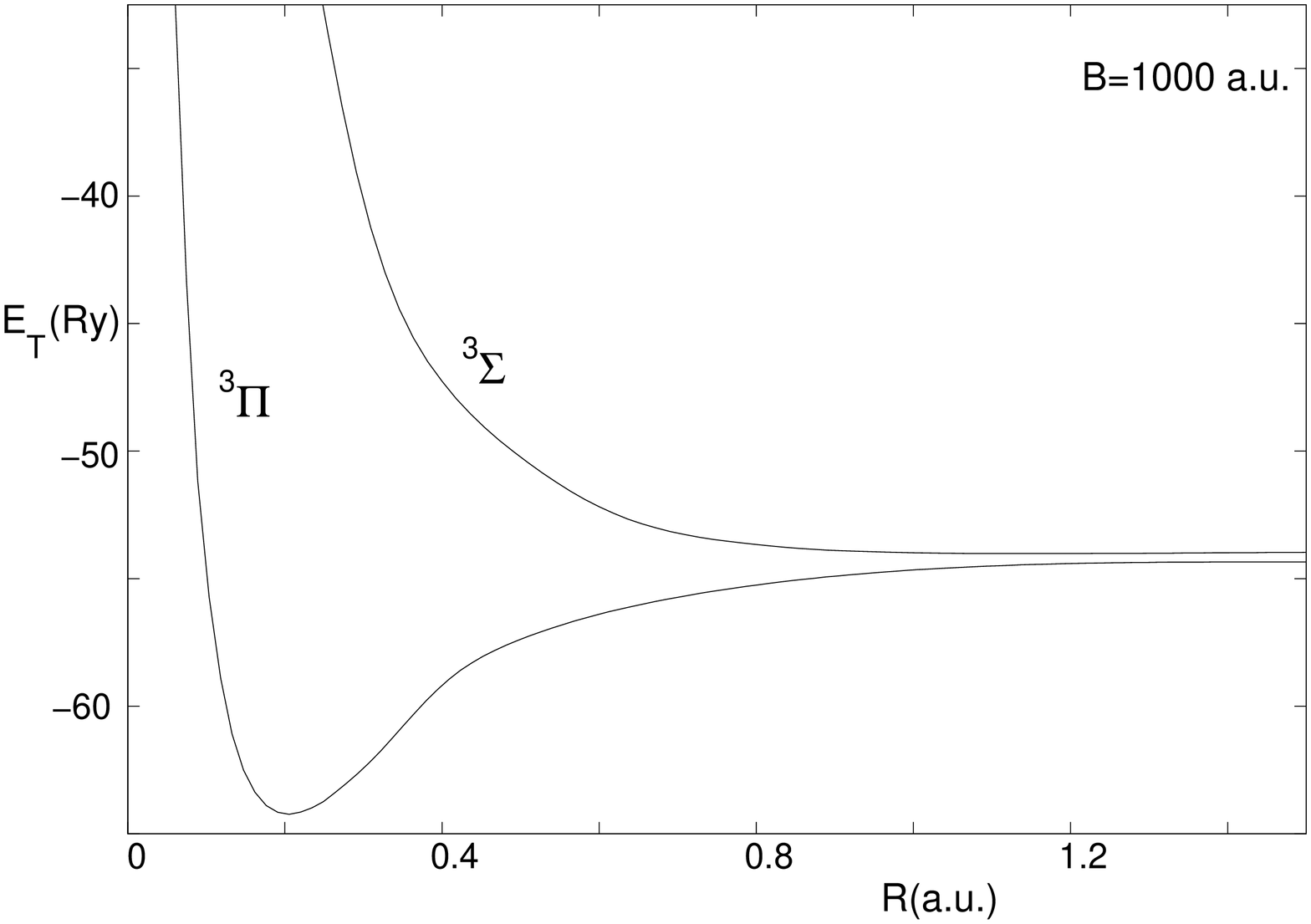}
    \caption{\label{fig:fig7} Total energy curves for triplet states
    $^3\Si$ and $^3\Pi$ for $B=1000$\,a.u.}
\end{center}
\end{figure}

\begin{figure}[tb]
\begin{center}
   \includegraphics*[width=3in,angle=-90]{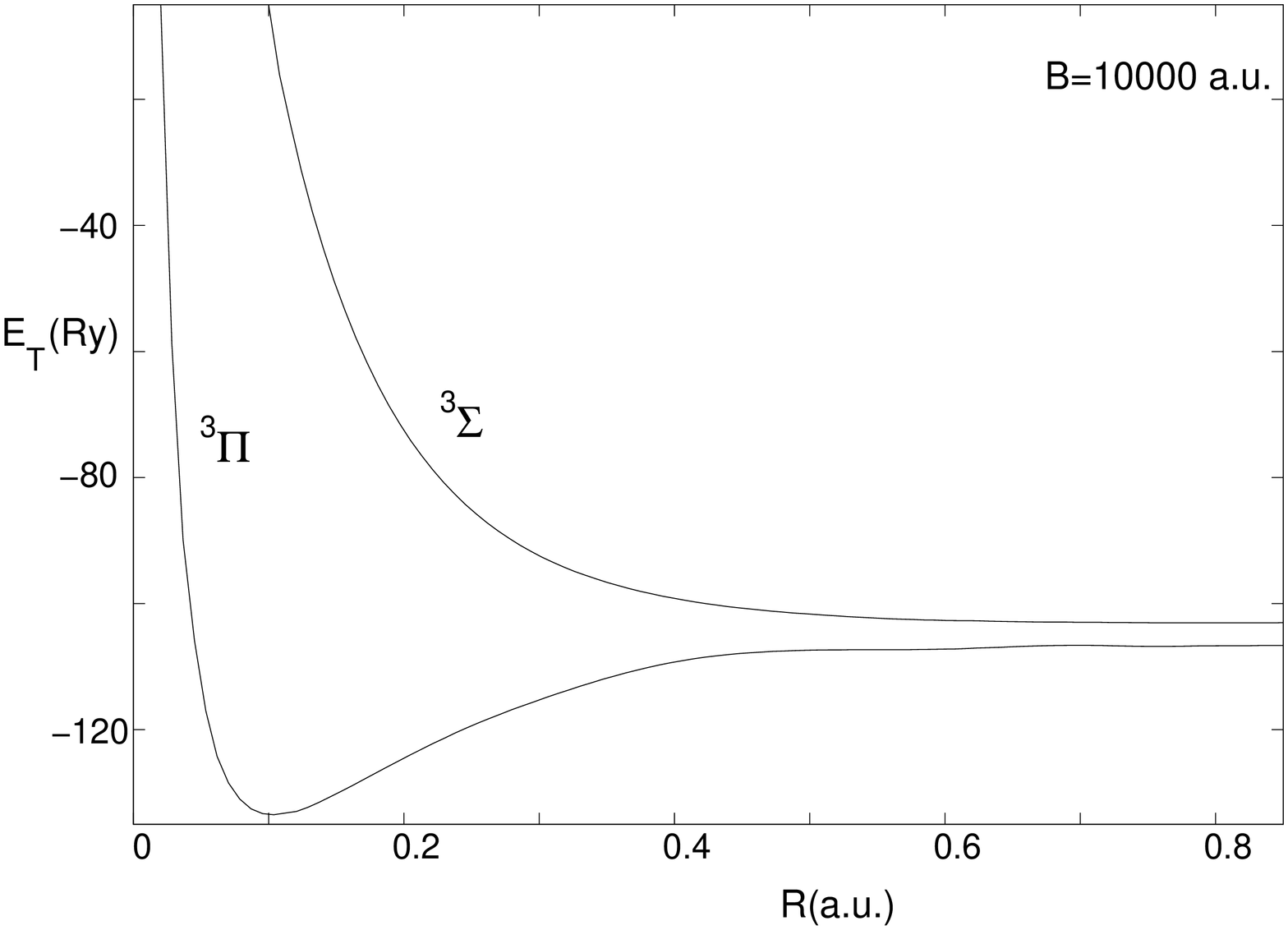}
    \caption{\label{fig:fig8} Total energy curves for triplet states
    ${}^3\Si$ and ${}^3\Pi$ for $B=10000$\,a.u.}
\end{center}
\end{figure}

\section{Conclusion}

The low-lying electronic states ${}^1\Si,{}^3\Si,{}^3\Pi,{}^3\De$ of
the Coulomb system $\al p e e$ in parallel configuration are studied
for $B=0-4.414\times 10^{13}$\,G using the variational method in the
Born-Oppenheimer approximation of the zero order. It is shown that
for all studied states and all magnetic fields the parallel
configuration is always optimal being stable with respect to small
inclinations. The state of the lowest total energy depends on the
magnetic field strength. It evolves from the spin-singlet ${}^1\Si$
state for small magnetic fields  $B \lesssim 0.5$\,a.u. to
spin-triplet $^3\Si$ (unbound state) for $B \gtrsim 0.5$\,a.u. and,
finally, for $B \gtrsim 15$\,a.u. to spin-triplet ${}^3\Pi$ state
(strongly bound state). This result shows that the $\rm{HeH}^+$
molecular ion exists as a bound state for magnetic fields $B
\lesssim 0.5$\,a.u. and $B \gtrsim 15$\,a.u. in a parallel
configuration and it is stable. In a domain $15 \lesssim B \lesssim
300\,$a.u. the state having the total energy next after the one of
the ground state is the unbound (repulsive) ${}^3 \Si$ state. For $B
\gtrsim 300$\,a.u. the ${}^3\Delta$ state is the lowest excited
state of the $\rm{HeH}^+$ molecular ion. As for $0.5 \lesssim B
\lesssim 15\,$a.u. the system $(\al p e e)$ likely exists in a form
of two separate atomic-type systems $\rm{He}^+(1s) + H(1s)$ with
total electron spin equal to one which are situated at very large
distance from each other. It is worth emphasizing that for $B\gtrsim
2000\,$a.u. both one- and two-electron ions $(He H)^{2+}$ and $(He
H)^{+}$ can exist. As for the ion $(He H)^{2+}$ it becomes stable at
$B\gtrsim 10000$\,a.u.

\begin{acknowledgments}
The authors are grateful to J.C. L\'opez Vieyra for numerous useful
discussions. This work was supported in part by FENOMEC and PAPIIT
grant {\bf IN121106} (Mexico).
\end{acknowledgments}

%%%%%%%%%%%%%%%%%%%%%%%%%%%%%%%%%%%
%       END OF FILE


\begin{thebibliography}{99}

\bibitem{Sanwal:2002}
         D.~Sanwal, G.G.~Pavlov, V.E.~Zavlin and M.A.~Teter,
%         ``Discovery of absorption features in the X-ray spectrum of an isolated neutron
%         star'',\\
         {\it ApJL \bf 574}, L61 (2002)\\
         (astro-ph/0206195)

\bibitem{PR}
           A.V.~Turbiner and J.C.~Lopez Vieyra, `One-electron Molecular Systems
           in a Strong Magnetic Field',
            {\it Phys.Repts. \bf 424}, 309-396 (2006)

\bibitem{HeH++}
           A.V.~Turbiner and J.C.~Lopez Vieyra,
           `The $(HeH)^{2+}$ and $He_2^{3+}$ exotic molecular ions can exist
           in a strong magnetic field',\\
%            Preprint ICN-UNAM 04-15, pp.7, December 2004\\
            {\it Int.Journ.Mod.Phys. \bf A} (2007) (in print)\\
                 (astro-th/0412399)

\bibitem{turbinerH2}
          A.V.~Turbiner, {\it Pis'ma ZhETF} {\bf 38}, 510(1983)

\bibitem{schmelcherH2}
          T. Detmer, P. Schmelcher, and L. S. Cederbaum, {\it Phys.Rev. \bf
          A57}, 1767 (1998)

\bibitem{Turbiner:2006H3+}
          A.V.~Turbiner, N.L.~Guevara and J.C.~Lopez Vieyra,
          `The $H_3^+$ molecular ion in a magnetic field: Linear
          parallel configuration',\\
            (physics/0606083)

\bibitem{Turbiner:2006He2++}
          A.V.~Turbiner and N.L.~Guevara,
          `The $\rm{He}_2^{2+}$ molecular ion can exist in a magnetic field'\\
            (astro-th/0610928)\\
          {\it Phys.Rev. \bf A 74}, 063419 (2006)

\bibitem{Turbiner:2006London}
             A.V.~Turbiner,
             `Molecular systems in a Strong Magnetic Field -
             how atomic - molecular physics in a strong magnetic field
             might looks like',\\
             Preprint ICN-UNAM 06-03, June 2006, pp.10\\
            {\it Astrophysics and Space Science} (2006)
             (in print)

\bibitem{beach}
          J.Y.~Beach,  {\it J.Chem.Phys. \bf 4}, 353 (1936)

\bibitem{anex}
          B.G.~Anex, {\it J.Chem.Phys. \bf 38}, 1651 (1963)\\
          L.~Wolniewics, {\it J.Chem.Phys. \bf 43}, 1087 (1965)\\
          C.~Urdaneta, A.~Largo-Cabrerizo, J.~Lievin et al,
%          G.C.~Lie and E.~Clementi,
          {\it J.Chem.Phys. \bf 88}, 2091 (1988)

\bibitem{3sigma}
          H.H.~Michels, {\it J.Chem.Phys. \bf 44}, 3834 (1966)\\
          T.A.~Green, H.H.~Michels and J.C.~Browne,
          {\it J.Chem.Phys. \bf 69}, 101 (1978)\\
          W.~Kolos, {\it Int.J.Quantum Chem., \bf 10}, 217 (1976)\\
          F.B.~Yousif, J.B.A.~Mitchell, M.~Rogelstad et al,
%  A.~Le~Paddelec, A.~Canosa and M.I.~Chibisov,
          {\it Phys.Rev. \bf A49}, 4610 (1994)

\bibitem{hogness}
          T.R.~Hogness and E.G.~Lunn,
          {\it Phys. Rev. \bf 26},  44 (1925)\\
          Z. Liu and P. B. Davies,
          {\it J.Chem.Phys. \bf 107}, 337 (1997)

%\bibitem{dalgarno}
%          W.~Roberge and A.~Dalgarno, {\it ApJ \bf 255}, 489
%          (1982)\\
%          B.~Zygelman, P.C.~Stancil, and A.~Dalgarno, {\it ApJ \bf 508}, 151
%          (1998)\\
%          D.~Galli and F.~Palla, {it Astron. \& Astrophys. \bf 335}, 403
%          (1998)\\
%          A.~Saenz, {\it Phys.Rev. \bf A67}, 33409(2003)

\bibitem{lepp}
          S.~Lepp, P.C.~Stancil and A.~Dalgarno,
          {\it J.Phys. \bf B35}, R37 (2002)

\bibitem{heyl}
          J.S.~Heyl and L.~Hernquist, {\it Phys.Rev \bf A58}, 3567 (1998)

\bibitem{turbinervar}
         A.V.~Turbiner, {\it Usp. Fiz. Nauk. \bf 144}, 35 (1984)\\
          {\it Sov.Phys. -- Uspekhi \bf 27}, 668 (1984)
          (English Translation)

\bibitem{schmelcherHe1}
           W.~Becken, P.~Schmelcher and  F.K.~Diakonos,
           {\it J.Phys. \bf B32}, 1557 (1999)

\bibitem{schmelcherHe}
           M.V.~Ivanov and  P.~Schmelcher,
           {\it Phys.Rev. \bf A61}, 022505 (2000)

\bibitem{Greennew}
           T.A.~Green, H.H.~Michels and J.C.~Browne,
           {\it J.Chem.Phys. \bf 64}, 3951 (1976)

\end{thebibliography}
\end{document}